\documentclass[twocolumn,showpacs,preprintnumbers,amsmath,amssymb]{revtex4}

\usepackage{graphicx}
\usepackage{dcolumn}
\usepackage{bm}
\usepackage{natbib}
\begin{document}

\preprint{Phys. Rev. Lett. 101, 085505 (2008)}

\title{Electronic Selection Rules Controlling Dislocation Glide in bcc Metals}
\author{Travis E. Jones$^\dagger$}
 \email{trjones@mines.edu}
\author{Mark E. Eberhart$^\dagger$}
 \email{meberhar@mines.edu}
\author{Dennis P. Clougherty$^{\dagger\ddagger}$}
\author{Chris Woodward $^{\natural}$}
\affiliation{$^\dagger$ Molecular Theory Group,
                         Colorado School of Mines,
                         Golden, Colorado 80401 }

\affiliation{$^\ddagger$ Department of Physics,
                   University of Vermont,
                   Burlington, Vermont 05405-0125 }

\affiliation{$^\natural$ Materials and Manufacturing Directorate,
                   Air Force Research Laboratories,
                   Wright Patterson Air Force Base, Dayton, Ohio 45433-7817 }
\date{\today}

\begin{abstract}
The validity of the structure-property relationships governing the deformation behavior of bcc metals was brought into question  with recent {\it ab initio} density functional studies of isolated screw dislocations in Mo and Ta.   These existing relationships were semi-classical in nature, having grown from atomistic investigations of the deformation properties of the  groups V and VI transition metals.  We find that the correct form for these structure-property relationships is fully quantum mechanical, involving the coupling of electronic states with the strain field at the core of long  $a/2\langle111\rangle$  screw dislocations.  
\end{abstract}

\maketitle


Single crystal bcc metals show remarkable variation in their plastic anisotropy and glide response.  For example, the anisotropy ratio of Ta is twice that of Mo, and under antitwinning stress, Ta is characterized by anomalous glide along (112) planes.  Based largely on the results of atomistic calculations \cite{Vitek, Duesbery1973, Xu, Duesbery1998}, these property variations were thought to originate from the structural differences of long screw-character $a/2\langle111\rangle$ dislocations, with the dislocation cores for Group V transition metals (Ta) being symmetric and those of Group VI metals (Mo), asymmetric. 

Beigi and Arias \cite{Beigi} were the first to question this assumption.  Using electronic structure Density Functional Theory (DFT) calculations to study of a closely spaced dislocation dipole arrays, they found qualitative evidence of symmetric strain fields around $a/2\langle111\rangle$ screw dislocations in Mo. A more realistic and quantitative DFT study \cite{woodward2001,woodward2002} employing a flexible boundary-condition also showed both Mo and Ta to be characterized by symmetric dislocation cores.  Further, the calculated anisotropy ratio agreed well with the limited experimental measurements, with a twinning anti-twinning asymmetry ratio of 2 in Mo and 4 in Ta.  More recent DFT investigations have shown that all of the group VI bcc transition metals are expected to produce symmetric cores \cite{woodward2005}.  

Despite the computational successes, the underlying structure(s) mediating the deformation properties of bcc metals remain elusive.  Here, we show that their variations in glide properties are due to differences in the symmetry imposed coupling between electronic states and applied strain.  The selection rules governing this coupling are mediated by the topology of the total charge density at the cores of $a/2\langle111\rangle$ screw dislocations. It is because the group VI metals have one more valence electron than do metals of group V, that their charge density topologies, and consequent properties, are also different.  (Though the variations in plastic anisotropy are also rooted in differences in electron count and charge density topology, this topic will be saved for a subsequent paper.)

The relationship between charge density topology and mechanical properties can be understood from  the Hohenberg-Kohn theorem, which asserts that a systemÕs ground-state properties are a consequence of its charge density, a scalar field denoted as $\rho(\vec{r})$ \cite{HK}.  Bader \cite{AIM} noted that the essence of a moleculeÕs structure must be contained within the topology of $\rho(\vec{r})$.  The topology of a scalar field is given in terms of its critical points (CPs), which are the zeros of the gradient of this field. There are four kinds of CP in a three-dimensional space: a local minimum, a local maximum and two kinds of saddle point. These CPs are denoted by an index, which is the number of positive curvatures minus the number of negative curvatures.  For example, a minimum CP has positive curvature in three orthogonal directions; therefore it is called a (3, 3) CP. The first number is simply the number of dimensions of the space, and the second number is the net number of positive curvatures. A maximum is denoted by (3, -3), since all three curvatures are negative. A saddle point with two of the three curvatures negative is denoted (3, -1), while the other saddle point is a (3, 1) CP.  For the purposes of this paper, only the (3, -3) and (3, -1) CPs need further consideration.

Through extensive studies of molecules and crystals, Bader and Zou \cite{Zou} and Bader \cite{AIM} showed that it was possible to correlate topological properties of the charge density with elements of molecular structure and bonding.  A maximum, a (3, -3) CP, is always found to coincide with the atomic nucleus, and so is called an atom CP. A bond path was defined to be the ridge of maximum charge density connecting two nuclei such that the density along this path is a maximum with respect to any neighboring path.  Bader showed that these ridges correlate with the locations of chemical bonds, leading one to associate the topologically unambiguous bond path with the sometimes subjective chemical bond.  The existence of a bond path is guaranteed by the presence of a (3, -1) CP between nuclei.  As such, this critical point is often referred to as a bond CP. 

\begin{figure}
\includegraphics{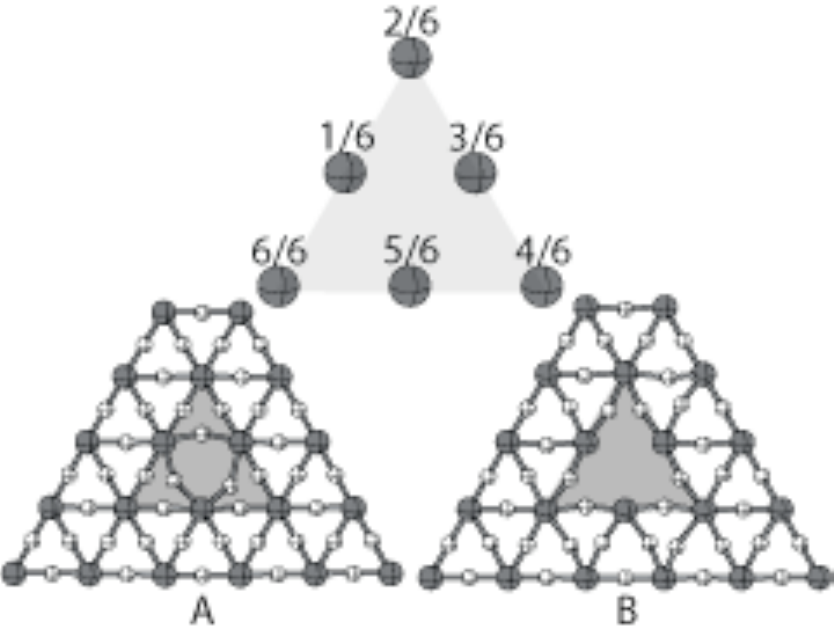} 
\caption{\label{fig:fig1}A projection onto the $(111)$ plane of stress free Mo (A) and Ta (B) cores.  Solid spheres give atom positions and open spheres are the locations of bond CPs.  Bond paths are shown with connecting lines. The core regions are shaded with triangles. Bonds paths are present across the Mo core, but absent in Ta. Outside the core region the topologies of the two metals are identical.  The insert (top) shows the fractional displacement of the atoms in the z direction (in and out of the page) for the atoms immediately adjacent to the core. Any full circuit around the dislocation results in a translation along z equal to one Burgers vector. }
\end{figure}

With a rigorous description of a bond path as a topological link, it is now possible to identify the bonds between atoms at dislocation cores.  Using the charge densities of Reference 5, the bulk topologies of both Ta and Mo were found to be those typical of bcc metals, with eight bond paths to first neighbors (there are no second or higher neighbor bond paths).  However, the topological structures of the two equilibrium dislocation cores are distinct (see Fig.~\ref{fig:fig1}).  Of importance is the recognition that bond paths cross the dislocation cores of Mo but not Ta.  Thus, while the strain fields of the screw dislocations in Mo and Ta are congruent, the connections between atoms are different.  The atoms at the dislocation core of Ta are deficient in two bonds, with only six, while Mo has the full complement of eight, as in the bulk. 

\begin{figure}
\includegraphics{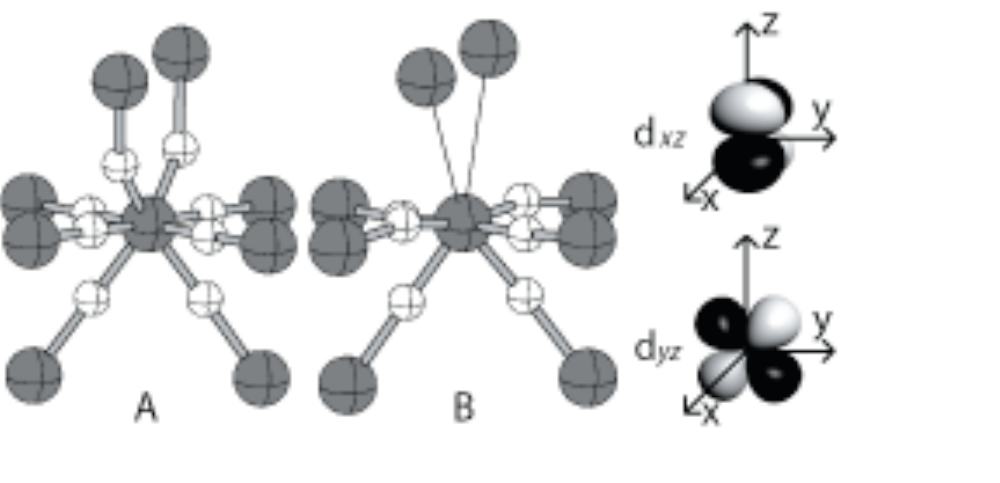} 
\caption{\label{fig:fig2} Structure around Mo (A) and Ta (B) core atoms corresponding to those labelled 1/6, 3/6, and 5/6 in Figure 1.  For Mo, the symmetry of the charge density about the central atom is nearly $D_{2d}$ - a rotation by 90 degrees about the z-axis followed by a reflection in the xy-plane will carry the BCP above the plane into those below the plane. Under this symmetry, the $d_{xz}$ orbitals, which participate in the bonding above the xy-plane, are nearly degenerate with the $d_{yz}$ atomic orbitals, which form the two bonds below the plane.  In Ta, the absence of BCPs above the plane reduces the symmetry of the charge density to $C_{2v}$, lifting the degeneracy between the $d_{xz}$ and $d_{yz}$ atomic orbitals.}
\end{figure}

Consider now the atoms labelled $1/6$, $3/6$, and $5/6$ in Fig.~\ref{fig:fig1}.  They are characterized by four bound atoms on an equatorial plane, with the remaining four near neighbor atoms situated on two perpendicular axial planes (see Fig.~\ref{fig:fig2}).  For the sake of clarity, we define a reference frame with an origin coincident with the $5/6$-atom.  The z-axis lies along a $\langle111\rangle$ direction and passes through the center of the dislocation.  This axis is normal to the $(110)$ equatorial plane containing the x and y axes of the local reference frame.  One may decompose the charge density into its contributions from all the atomic orbitals and find that with respect to the reference frame, the $d_{xy}$ orbital on the central metal atom (Mo or Ta at the $5/6$ position) contribute density to the bond paths in the equatorial plane.  In the case of Mo, the $d_{xz}$ and $d_{yz}$ orbitals contribute density to the bond paths above and below this plane.  While for Ta, without bond paths above the plane, there is almost no contribution to the total charge density from the $d_{xz}$ orbital of the central metal atom.  

In Ta, symmetry breaking provides the driving force for occupying the $d_{yz}$ at the expense of the $d_{xz}$.  For Mo, the symmetry of the charge density about the $5/6$ atom is nearly $D_{2d}$ - by virtue of an improper 4-fold rotation about the z-axis (see caption Fig.~\ref{fig:fig2}).  Under this symmetry, the $d_{xz}$ and $d_{yz}$ orbitals will be nearly degenerate.  The projected density of states for the Mo dislocation core confirms this fact, with a nearly full d-band derived from the $d_{xz}$ and $d_{yz}$ orbitals of the $1/6$, $3/6$, and $5/6$ type atoms.  In the case of Ta, this band would be half full, but is split into an occupied and unoccupied portion.  The disappearance of the bond points above the plane lowers the symmetry to nearly $C_{2v}$ and splits the band. The key here is that in Mo the $d_{xy}$, $d_{xz}$, and $d_{yz}$ orbitals are split into two groups, a degenerate pair ($d_{xz}$ and $d_{yz}$) and an orbital singlet ($d_{xy}$), while in Ta they are split into three nondegenerate groups.  

Consider now the charge redistribution that must accompany dislocation motion. In the bulk where the local coordination is cubic; the $d_{xy}$, $d_{xz}$, and $d_{yz}$ atomic orbitals will be degenerate.  As the dislocation moves, and the atoms at the core rearrange to form bulk, degeneracy is achieved through the flow of electrons between them.  In turn, this charge redistribution is permitted through the ÒcouplingÓ of the charge by the perturbation acting on the dislocation, in this case, the applied strain.
The quantum mechanical laws governing this coupling can be reduced to their principal factors by assuming the applied strain rate is slow, with the atoms of the dislocation core moving on an adiabatic potential surface.  Then, we can write the strain perturbation as a sum of two static parts: a term that depends only on the strain and one that depends on its spatial rate of change, i.e. 

\begin{equation}
\varepsilon(\vec r+d\vec r)\approx \varepsilon(\vec r)+d\vec r \cdot \nabla\varepsilon(\vec r)
\end{equation}
where $\varepsilon$ is the strain.  The second term on the right is simply the rate of change of the strain in the direction $d\vec r$. This term will be zero in a material with uniform elastic properties.  In a material with varying elastic properties, however, it will change most rapidly along the directions in which bonds are being made and broken due to the strain at $r$-Ðthe first term on the right. Thus, each of these terms can couple orbitals and facilitate charge redistribution; their importance in this process, however, will be determined by their relative magnitudes. 

The results summarized in Ref.~5 confirm that for both Ta and Mo dislocations moving under a twinning stress the magnitudes of the two terms in equation (1) are comparable, for as the strain is applied, new bonds form and the dislocation moves at a small Peierls strain.  On the other hand, in the anti-twinning sense, bond breaking and making occurs late in the reaction coordinate (large Peierls strain) and only after significant atomic rearrangement.  Hence, for anti-twinning, the initial response of the dislocations is dominated by the applied strain.  And, whereas the quantum mechanical constraints imposed on the charge redistribution are difficult to predict when $d\vec r \cdot \nabla\varepsilon(\vec r)$ is large, when it is small or vanishes (anti-twinning stresses), we can call upon the principles of symmetry to predict the atomic motions that couple orbitals and permit charge flow.  (See supplementary material)

Beginning with Mo, charge flow between the $d_{xz}$-$d_{yz}$ pair and the $d_{xy}$ orbital  is only permitted when atoms move in response to a shear stress in the $(110)$ plane.  On the other hand, in Ta, with three singly degenerate orbitals, there are two coupling strains.  The empty $d_{xz}$ orbital is coupled to the occupied $d_{yz}$ orbital by shear strains lying in a plane perpendicular to the  $(110)$ plane.  Shear strains in the $(110)$ plane couple the remaining orbitals--$d_{xy}$ to $d_{xz}$ and $d_{xy}$ to $d_{yz}$. The constraints imposed by these coupling rules, known formally as selection rules, are seen as Ta and Mo deform under an anti-twinning stress.  

As a rule, dislocations in metals move along close-packed planes.  Thus, in bcc metals, dislocations are expected to move along $(110)$ planes, which is the case for Mo.  For example, when a pure shear stress is applied in the $\langle111\rangle$ anti-twinning sense, dislocations in Mo move as expected (along the $(110)$ plane of maximum resolved shear stress).  In Ta, however, the dislocation responds to this shear stress on the $(112)$ plane.  Only after significant charge redistribution can bonds begin to form, at which point the dislocation propagates in response to the gradient terms of Equation (1).  Thus, the driving force for the motion is the formation of bonds using tantalum's empty $d_{xz}$ orbital.  The charge to form these bonds comes from electrons in the Ta $d_{yz}$ orbital.  This charge flow is not allowed if the dislocation is confined to the $(110)$ plane but is possible by coupling orbitals through the strain field in the $(112)$ plane.  The observed motion of dislocations in Ta under anti-twinning stress is consistent with the symmetry imposed selection rules. Ultimately, these differences stem from the fact that Ta has one fewer valence electron than does Mo.

With an electronic mechanism for dislocation motion in hand, the effects of dilute substitutional impurities to the mechanical properties of non-magnetic bcc metals are predictable.  Substitutional alloying of elements to the left of the Group V metals will deplete the d-band, yielding an alloy with properties more like those of Ta.  Conversely, those to the right will further fill the d-band, producing properties more like Mo.  This result is entirely reasonable based on purely empirical arguments.  However, it is not the conclusion, but the method by which it was deduced that is of most importance.  Combining density functional theory with the topological model of molecular bonding, we were able to uncover structure-property relationships that have eluded others.  The procedure used here should be equally applicable to problems of fracture and deformation where empirical findings do not provide guidance.  In such cases, this approach may shed light on unrecognized facets of alloy theory.

\subsubsection*{Supporting material}

With respect to the $D_{2d}$ point group, the $d_{xz}$ and $d_{yz}$ atomic orbitals transform as the doubly degenerate $E$ representation, while the $d_{xy}$ orbital transforms as the irreducible representation $B_2$.  If we take $\Gamma_v$ to be the irreducible representation of the strains that couple these two, then the fully symmetric representation, $A_1$, must be contained in the direct product of $E \times \Gamma_v \times B_2$.  From which one may show that  $\Gamma_v$ transforms as $E$, which possess the same symmetry as shear strains in the $(110)$ plane.  
With respect to the $C_{2v}$ point group, $d_{yz}$ reduces as $B_2$, $d_{xz}$ as $B_1$ and $d_{xy}$ as $A_2$.  Knowing that $B_2 \times B_1 \times A_2 = A_1$, it can be shown that: $d_{xz}$ is coupled to $d_{yz}$ by an $A_2$ strain (shear couple applied in a plane normal to the z axis), $d_{xy}$ is coupled to $d_{yz}$ by a $B_1$ strain (shear normal to the y axis), and $d_{xy}$ is coupled to $d_{xz}$ by a $B_2$ strain (shear couple normal to the y axis). $B_1$ and $B_2$ form a symmetry basis for the full set of shear strains in the $(110)$ plane of Fig.~\ref{fig:fig2}.

\begin{acknowledgments}   
We are grateful to the Defense Advanced Projects Agency and the Office of Naval Research for their support of this research.
\end{acknowledgments}

\bibliography{Refs3}

\end{document}